# STUDENTS LEARNING CENTER STRATEGY
# BASED ON E-LEARNING AND BLOGS

**Leon Andretti Abdillah**
Information Systems Study Program, Computer Science Faculty, Bina Darma University
Jl. Jenderal Ahmad Yani No.12, Plaju, Palembang, 30264
Email: leon.abdillah@yahoo.com

**Abstracts**
*Education is the main infrastructure for promoting a nation and increasing their competitiveness in globalization era that involves the use of information technology (IT). This paper has goal to expand the alternative for learning strategy based on IT for final year students that will conduct a research for their final report. The final year students are expected have the ability to learn independently to manage the needness of their learning supply. In this paper, the author will discuss how to use the e-learning media and blogs to manage the independently learning environment or self-learning. E-learning offers the flexibilities in term of time and place in supporting the learning activities as well as a media for faculty/lecturer to disseminate learning materials. While the blog has been a free medium for the publication of variety contents, including academic contents. In this papaer, author suggests some the institutions regulations, lecturers roles, and students participations are involve in set the learning environment based on e-learning and blog. These activities able to support students in learning independently or self-learning.*

**Abstrak**
*Pendidikan merupakan infrastruktur utama untuk memromosikan suatu bangsa dan meningkatkan tingkat kompetitif mereka di era globalisasi yang melibatkan penggunaan Teknolog Informasi (TI). Paper ini memiliki tujuan untuk memperluas alternatif strategi sentra pembelajaran mahasiswa (student learning centre) berdasarkan TI (IT-based learning). Mahasiswa diharapkan memiliki kemampuan untuk belajar mandiri dalam mengelola kebutuhan belajar mereka. Pada paper ini, peulis akan mendiskusikan bagaimana menggunakan media "e-learning" dan "blogs" untuk mengelola lingkungan belajar independen atau "self-learning". E-learning menawarkan fleksibilitas pada hal waktu dan tempat dalam mendukung kegiatan pebelajaran dan juga sebagai media bagi fakultas/dosen untuk mendistribusikan materi-maeri pembelajaran. Sementara blogs telah menjadi media gratis untuk publikasi beragam konten, termasuk konten akademik. Penulis menyarankan skema yang terdiri atas; 1) regulasi institusi, 2) peranan dosen, 3) dan partisipasi mahasiswa yang terlibat dalam setting lingkungan pembelajaran berbasis e-learning dan blogs. Kegiatan-kegiatan tersebut dapat mendukung para mahasiswa dalam belajar secara independen.*

## INTRODUCTION

Education is a primary necessity early adulthood should be felt by the whole community (Abdillah and Emigawaty, 2009). Education also the vital variable for the developing a nation. A nation could move forward compare to others countries because the spread of knowledge on the country. Knowledge developed in line with the needs and challenges for the better future. The most responsible institutions for spreading the knowledge is education institutions. Era of globalization, the requirement with the development of information technology (IT) makes the world of education must adopt and involve IT in the learning process. In fact, the current students were born in digital era where technology has become their close friend from first supporting environment for them. The main focus of this paper is IT based education.

This paper offers the strategy scheme of student learning centre by making use of e-learning and blog. E-learning has changed the phenomenon of learning become placeless, boarderless and timeless. Major benefits of e-learning included the ease of access to resources, and the provision of central area for students to access to find information or comprehensive resources pertaining to each module (Concannon et al., 2005). Students able to ask the problem related to the course material any time, and also triger student self-learning (Djajalaksana, 2011). Several authors have found that the most efficient teaching model is a blended approach, which combines self-paced learning, live e-learning, and face-to-face classroom learning (Alonso et al., 2005).

The software used for e-learning is Moodle (Moodle, 2013). Moodle is a Course Management System (CMS), also known as a Learning Management System (LMS) or Virtual Learning Environment (VLE). It's a free web application that educators can use to create effective online learning sites. The implementation e-learning, will achieve following objectives (Hasibuan and Santoso, 2005):

- Give a priority to development of regional based higher education by developing focal points.
- Increase the access and equality of education.
- Improve and disseminate relevancy and quality of education.
- Improve research activities, both in terms of quantity and quality.

On the other hand, Weblog (blog) helps as the media for disseminating the knowledge through internet. Blog is design as personal web, where the author of the blog commonly recognize as blogger. Blog become one of the most popular icons in internet application nowadays. It offers many features that owner able to customize them. The owner has dominant authorship to populate it with many useful contents. The blog's owner able to use blog for online diary, business promotion, personal webpage, even for education purposes. The features of blog could more effectively visualize and represent student's learning, as well as communicate with others. Blog also proves to be an effective tool that enables students learning in an elearning environment (Lin et al., 2006). Some of lecturers use it to share their class meeting material to their students. Others use it to share their documents or presentations of research and workshop (Sanjaya and Pramsane, 2008), blogging offers particularly useful opportunities for learner-centered feedback and dialogue (Glogoff, 2005), and the author/writer it self as a lecturer also use blogs to document the past supervised students with their topics, and other professional servies. In this paper, author focus on the content for distance learning or e-learning and uses Wordpress for the experiment because it is popular, easy to use, and provide many features for elearning purposes. In terms of e-learning, there are two important features of blogs (Jung et al., 2006), as follows: 1) Personal content management (e.g., presentation files, examples, and webpages), and 2) Information propagation by social activities. (e.g., questioning and replying).

The combination of these two media provides center environment for student learning. E-learning is the activity where the learning atmosphere involved lecturers and students. Lecturers are able to disseminate the knowledge (learning materials) to the students. On the other hand, students also able to participate in learning scenario, submit their assignments, give comments, or discuss with their colleagues.

The rest of this paper will discuss methods (section 2), discussions (section 3), and conclusions (section 4).

**METHOD**

A total 264 students (fifth and seventh undergraduate students) participated in the observations. The subject in main core subject is computer science/information systems field. The goal of this paper is to set the strategies approach in manage student learning environment based on IT. The subject course is in computer science/information systems field. Author uses free e-learning software (moodle), weblog application (wordpress), mobile operating systems for Apple (iOS), and blog application for iPhone (Wordpress for iPhone).

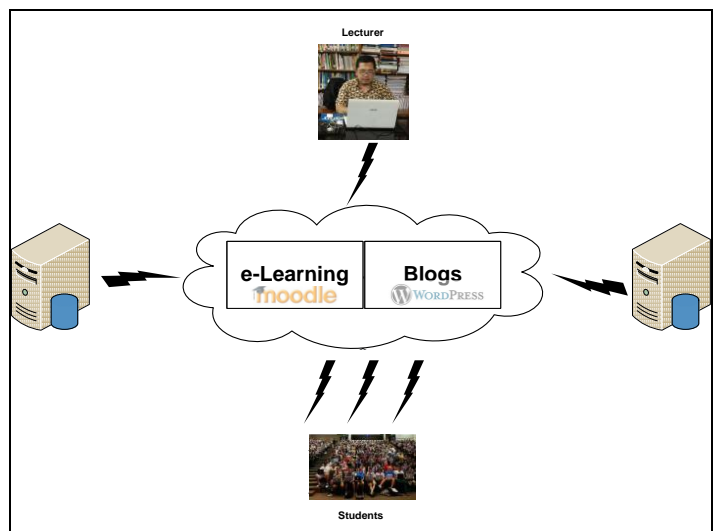
**Figure 1.** Blended e-Learning scheme

This paper discuss the blended e-learning strategy involves lecturer, e-learning, blogs, students, and server (figure 4). Based on the data collected and observations of the activities, the authors will discuss about several strategies to conduct blended e-learning. The author will cluster the duscusions into three groups, institution regulations, lecturer roles, and students participations.

Lecturer combines severals conditions to conduct the blended learning involved class meeting, e-learning, and group presentations. The total general meeting are 14 meetings. Two of them are e-learning meetings, two meetings use for exams (9 and 14), one meeting for group presentations (13). Every class meeting will involved 31 – 55 students. During the semester, a subjet consist of six main activities, namely: Daily Attendances, Daily Test (5), Mid Test (9), Weekly Reports (2-12), Presentation (13), and Final Test (14). For the illustration, please have a look the figure 1.

**DISCUSSIONS**
*Institution regulations*

Abdillah (Abdillah et al., 2007) has declared that lecturers's performance are influenced by compensations. To motivate lecturers in providing high quality academic knowledge and actively involved in e-learning processes, institution needs to lunch popular e-learning competitions program. This program will trigers lecturers to show their performance in conduct the learning process through e-learning process. Institution should supply the most active lecturers with interesting compensations and high credit for their performance.

Another aspect needs to be consider is the lecturers background. We should reliaze that lecturer from computer science background are more familiar with IT. It means, institusion needs to differ the competition based on lecturers background. The last aspect is the criteria for the e-learning competition should involve various activities (figure 2).

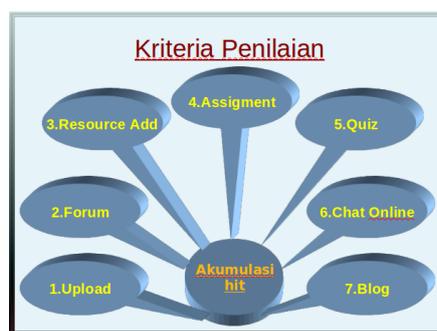
**Figure 2.** e-Learning competition criteria

*Preparation*

In the first stage, lecturer need to install an iOS App for Wordpress (figure 3), prepare blog address, and set up the array of serial meetings through one semester (approx. 3 – 4 months). The course material for each meeting will be supplied before the meeting in the most popular article format, pdf (Abdillah, 2012) and/or ppt forms. Students need to download and read every single material carefully. Lecturer need to upload the course materials through e-Learning and blog.

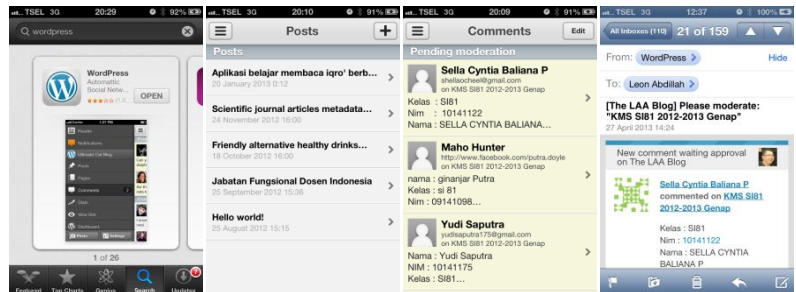

**Figure 3.** iOS App for Wordpress

*Learning process*

In the second stage, lecturer need to manage the 14 to 15 weeks/meetings (figure 4). will explain and discuss every material in the class meeting. At the first meeting, time slot will be allocated for course explainations, course rules, review for the prerequested subjects (optional), and introducing the current materials, grade systems, references, and chapter 1.

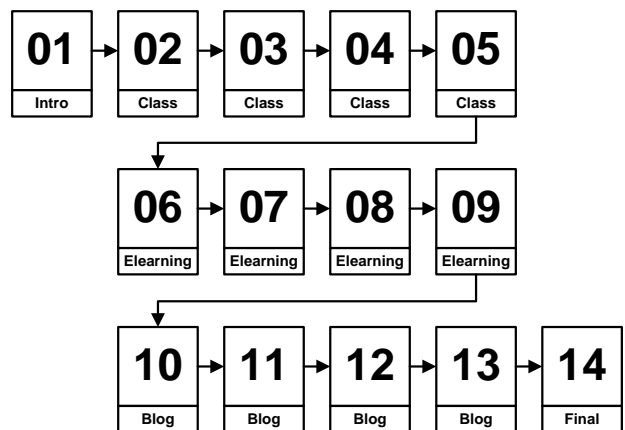

**Figure 4.** Meeting flow scenario

The first month, lecturer focus on face to face meeting in the class. While giving the knowledge in oral traditional class meeting, this activity also involve set up group discussions in every class, try to use e-learning facilities, and manage the link from every students blog address. The second month, lecturer no long focus on the claa meting, instead more active in e-learning facilities, such us a) forum, b) give feedbacks to any submitted assignments, c) chat with online participant students, etc. The third month, lecturer move to share the knowledge through blogs.

*Focus groups*

Start from the begining, each class will be set up by several groups. Each group consist of 3-7 students depend on total population in the class. The main goal of this strategy is to organize the behaviour of student learning in team work. Each group will ask to choose a specific theme for their project, and need to share the duties with their members. After they finish with the weekly job, every student need to submit their work through e-learning and blog in period of times. Even most of the students in their group submit the assignment, but if a student does not submit the assignment then that student will not gain the credit. This strategy will trigers students for team work.

This activity will encourage every student for weekly discussion in their group. For a semester, this activity will produce five e-learning submitted assignments and five published blogs assignments.

*Exam and Evaluation*

Overall there are only three exams through the semester, daily exam, mid exam, and final exam. In order to engage the students with e-learning materials and blog publications, lecturer needs to link the qustions in the exam with e-learning and blog. But, according to lecturers experiences, the exam should not conduct in e-learning, because many of the students answer the questions in very short of time. In this research, author combine the exams with some assignments through e-learning and blogs publication. It means if a student only involve with classical exam without active in e-learning and blog then their marks could not reach optimum score.

**CONCLUSIONS**

IT offers many benefits to learning systems. It has changed the way of learning styles and approaches. IT dominantly uses to support IT based university. Some technology to support IT learning environment are e-Learning and blogs. Author focus on this two technologies to set the strategy for student learning center.

Based on the experiences and observations, author classify the point of view of this paper based on: 1) institutions regulations supports, 2) lecturers roles, 3) e-learning technology, and 4) participants of students.

First of all, the institution need to support the e-learning environment in terms of infrastructures, interesting compensations for lecturers, and the supporting resources (internet connections, staff, atc.). Institution also need to consider the lecturers backgrounds, and give appropriate training to all lectures.

E-learning changes the traditional learning model become visual and faceless. The lecturers able to supply the subject materials for the learners (students), control the updateness of the subject, explore the activities of each the participants such us attendance online, assignment submissions, etc. Sometimes, lecturers also able to set chat online with the participants, announce several information, and give valuable feedbacks to the learners. In this paper, author also use smartphone to monitor the students activities in blog.

In this paper, author combine the e-learning environment with blogs application. Blog able to enrich the facilities of e-learning. Blogs offer the free service to link students with social media and their colleages. Students will engage in fun environment where they have media to express their academic resipotories and connections. This strategy will motivate students to keep participate to build their e-learning experiences, involve in distance learning and see their work globally. To set the conductive learning experiences, lecturers need to set group study in every class.

For the next research, authors still interested to work in e-learning environment but with some extentions involving the benefits of social media powers and also combine it with various of widgets. The next research involve more applications and enrich the learning strategy and environment.